# Optimization of Pressure Management Strategies for Geological CO2 Sequestration Using Surrogate Model-based Reinforcement Learning

Jungang Chen

jungangc@tamu.edu

*Harold Vance Department of Petroleum Engineering, College of Engineering,*

*Texas A&M University, College Station, Texas, USA*

Eduardo Gildin, Ph.D.

*Professor, Harold Vance Department of Petroleum Engineering, College of Engineering,*

*Texas A&M University, College Station, Texas, USA*

Georgy Kompantsev

*Harold Vance Department of Petroleum Engineering, College of Engineering,*

*Texas A&M University, College Station, Texas, USA*

# Optimization of Pressure Management Strategies for Geological CO2 Sequestration Using Surrogate Model-based Reinforcement Learning

**ABSTRACT**

Injecting greenhouse gas (e.g. CO2) into deep underground reservoirs for permanent storage can inadvertently lead to fault reactivation, caprock fracturing and greenhouse gas leakage when the injection-induced stress exceeds the critical threshold. It is essential to monitor the evolution of pressure and the movement of the CO2 plume closely during the injection to allow for timely remediation actions or rapid adjustments to the storage design. Extraction of pre-existing fluids at various stages of the injection process, referred as pressure management, can mitigate associated risks and lessen environmental impact. However, identifying optimal pressure management strategies typically requires thousands of simulations, making the process computationally prohibitive. This paper introduces a novel surrogate model-based reinforcement learning method for devising optimal pressure management strategies for geological CO2 sequestration efficiently. Our approach comprises of two steps. Firstly, a surrogate model is developed through the embed to control method, which employs an encoder-transition-decoder structure to learn dynamics in a latent or reduced space. Leveraging this proxy model, reinforcement learning is utilized to find an optimal strategy that maximizes economic benefits while satisfying various control constraints. The reinforcement learning agent receives the latent state representation and immediate reward tailored for CO2 sequestration and choose real-time controls which are subject to predefined engineering constraints in order to maximize the long-term cumulative rewards. To demonstrate its effectiveness, this framework is applied to a compositional simulation model where CO2 is injected into saline aquifer. The results reveal that our surrogate model-based reinforcement learning approach significantly optimizes CO2 sequestration strategies, leading to notable economic gains compared to baseline scenarios.

## 1 Introduction

The challenge of mitigating climate change has led to the exploration of various carbon capture and storage (CCS) techniques, among which geological CO2 sequestration presents a feasible, reliable and scalable solution. This process involves the injection of greenhouse gases, primarily carbon dioxide ($CO_2$), into deep geological formations for long-term storage. Despite its potential, one of the primary concerns in geological CO2 sequestration is the risk of inducing mechanical instabilities, such as fault reactivation and caprock fracturing [1, 2]. These events can occur when the stress induced by CO2 injection surpasses critical thresholds, potentially leading to unintended greenhouse gas leakage. Therefore, managing the pressure to avoid surpassing these critical stress thresholds is crucial [3, 4]. Traditionally, pressure management strategies have involved the extraction of pre-existing fluids during various stages of the CO2 injection process. While this approach can effectively mitigate risks associated with over-pressurization and reduce the environmental impact, it requires careful planning and optimization. Identifying optimal pressure management strategies often necessitates extensive computational modeling and simulation, which can be both time-consuming and resource-intensive.

Employing reduced-order models (ROMs), also referred as surrogate models and proxy models, for planning and optimization has introduced a computationally efficient way to address aforementioned challenges. Previous studies [5, 6] underscore the application of ROMs for well controls optimization, focusing on their integration with gradient-based and derivative-free optimization algorithms. A study [7] investigates the deployment of a ROM named Proper Orthogonal Decomposition with Trajectory Piecewise Linearization (POD-TPWL) for CO2 sequestration, aiming to minimize the CO2 leakage into the caprock at the end of storage period via a derivative-free algorithm known as Mesh Adaptive Direct Search. Notably, the emergence of deep learning-based ROMs has been recognized for offering superior alternatives to their traditional counterparts [8, 9, 10], showcasing enhanced potential in simplifying and expediting the optimization process. [11] introduces deep learning-based surrogate model for subsurface flow, highlighting its capability to handle much wider range of control specifications compared to traditional ROMs because the deep learning-based ROMs can generalize better. [12] extends the framework by incorporating well outputs data formulation in the latent space and improves the prediction performance. However, their potential for application to CO2 sequestration and their capability to couple with optimization framework have not been fully explored.

Reinforcement learning (RL) has been recognized for its capability of planning and optimization problems. Depending on whether the model of the environment is accessible or not, RL can be divided into two categories, model-free reinforcement learning and model-based reinforcement learning. Model-free RL methods make decisions based solely on the rewards received from previous actions, without any understanding or model of the environment's dynamics. This approach, while simpler and more straightforward, focuses on the direct association between actions and outcomes. Model-free methods can be less efficient because they need to explore and learn from each possible state-action pair to make accurate predictions, which can be impractical in complex environments. Model-based reinforcement learning, on the other hand, involves learning a model of the environment's dynamics, including state transitions and rewards, to make decisions. Model-based RL can be more sample-efficient, however, it may suffer from curse of dimensionality for high-dimensional states. Surrogate model-based reinforcement learning aims to bridge the gap by using approximate or proxy models of the environment. These models are simpler and less computationally demanding than 'true' models but still capture enough of the environment's dynamics to allow for effective planning and decision-making. Notable contributions in this area include references [13, 14, 15, 16].

In this work, we introduce a surrogate model-based reinforcement learning framework tailored for devising optimal pressure management strategies for geological CO2 sequestration. Our approach consists of two primary components: Firstly, we employ the embed to control method, harnessing an encoder-transition-decoder structure to learn the latent dynamics inherent in CO2 injection process. The surrogate model significantly reduces the computational complexity by compressing the high-dimensional states into a lower-dimensional latent representation. Building upon the surrogate model, we use a model-free reinforcement learning algorithm called soft actor-critic (SAC) to derive strategies that maximize economic returns while adhering to various control constraints. This approach can lead to more efficient learning by combining the sample efficiency of model-based methods with the simplicity and computational tractability of model-free methods.

This paper is organized as follows: Firstly, we describe the fundamental compositional equations governing CO2 storage process and introduce its reduced-order modeling and optimization formulations. Following that, we present the framework of surrogate model-based reinforcement learning for the control optimization in CO2 sequestration scenarios. Conclusively, we underscore the efficacy of our approach through a case study involving CO2 injection into a saline aquifer, illustrating the practical benefits and the potential of our method in enhancing the environmental and economic outcomes of geological CO2 sequestration efforts.

## 2  Problem Description

### 2.1  Governing Equations

In this study, we consider a compositional formulation for the gas-water system, where mass conservation of each given component $i$ can be expressed as follows,

$$\frac{\partial}{\partial t}(\phi(\rho_w S_w c_{i,w} + \rho_g S_g c_{i,g})) + \nabla \cdot (c_{i,w}\rho_w \boldsymbol{v}_w + c_{i,g}\rho_g \boldsymbol{v}_g) = \sum_{j=1}^{n}(c_{i,w}\rho_w q_w^j + c_{i,g}\rho_g q_g^j)$$

(1)

Where $\frac{\partial}{\partial t}$ is the time differentiation, $\phi$ stands for porosity, $\rho_\alpha$ and $S_\alpha$ represents phase density and saturation of phase $\alpha = w, g$ respectively, $c_{i,\alpha}$ denotes mole fraction of component $i$ in phase $\alpha$. Within the context of a CO2-water system, we focus on CO2 and water as the primary components. $\boldsymbol{v}_\alpha$ is the volumetric velocity of phase $\alpha$ and $q_\alpha^j$ denotes the source/sink term for phase $\alpha$ and well $j$. $\boldsymbol{v}_\alpha$ is given by Darcy's law, which reads

$$\boldsymbol{v}_\alpha = -\boldsymbol{k}\frac{k_{r\alpha}}{\mu_\alpha}(\nabla p_\alpha + \rho_\alpha g \nabla D)$$

(2)

Here $\boldsymbol{k}$ refers to the total permeability; $k_{r\alpha}, \mu_\alpha, p_\alpha$ and $\rho_\alpha$ are the relative permeability, viscosity, pressure and density of $\alpha$-phase, respectively. $g$ denotes the gravitational constant. This set of compositional equations can be translated into a state-space framework, where reduced-order techniques can be used. [7, 17, 18] applies molar formulation and identifies two primary system variables of the system: the pressure $p$ and the overall mole fraction of either CO2 or water component, given by $z_{CO_2} = S_w c_{CO_2,w} + S_g c_{CO_2,g}$ or $z_w = S_w c_{w,w} + S_g c_{w,g}$. Note that $z_{CO_2} = 1 - z_w$.

Adhering to the system-control approach and assuming constant timestep size, the above equations can be transformed into the following discrete-time form:

$$x_{t+1} = f(x_t, u_t)$$
$$y_{t+1} = g(x_{t+1}, u_t)$$

(3)

Where $x_{t+1}$ and $y_{t+1}$ signifies system's state and outputs of next timestep $t+1$, respectively. $u_t$ is the well control applied at time $t$. By applying Taylor's expansion, the non-linear equations can be rewritten to a locally linear version, given by

$$x_{t+1} = A(x_t)x_t + B(x_t)u_t$$

$$y_{t+1} = C(x_t)x_{t+1} + D(x_t)u_t$$

(4)

Where $A(x_t) = \frac{\partial f}{\partial x_t}$, $B(x_t) = \frac{\partial f}{\partial u_t}$, $C(x_t) = \frac{\partial g}{\partial x_{t+1}}$ and $D(x_t) = \frac{\partial g}{\partial u_t}$.

In CO2 storage operations, specifically for CO2-water system, state variables $x_t$ entails pressure and overall mole fraction of CO2 component, represented as $x_t = \begin{bmatrix} p_t \\ z_{CO_2,t} \end{bmatrix} \in R^{2*N_b}$. The well controls $u_t$ and the system outputs $y_t$ are defined as $u_t = \begin{bmatrix} BHP_t \\ q_t^{inj} \end{bmatrix} \in R^{N_{prod}+N_{inj}}$ and $y_t = \begin{bmatrix} q_{t,w} \\ q_{t,g} \\ P_{wf,t} \end{bmatrix} \in R^{2*N_{prod}+N_{inj}}$, where $N_b$, $N_{prod}$, $N_{inj}$ represents number of grid blocks, number of production wells and injection wells, respectively.

Traditional reduced-order modeling (ROM) techniques, such as those using Proper Orthogonal Decomposition (POD), seek to project the original high-dimensional space $x$ to a lower-dimensional latent state space $z$ using the POD basis $\Phi$ derived from collected data snapshots [19, 20, 21, 22], e.g. $z = \Phi x$. Despite the effectiveness, their reliance on direct source code access could hinder their practical applications. Alternatively, deep learning-based ROMs offer a non-intrusive approach that can seamlessly integrate with existing commercial simulators. Furthermore, traditional ROMs assume that the dynamics of the system can be accurately captured in a linear subspace of the full state space. This assumption can limit its effectiveness in handling systems with highly nonlinear behavior. On the other hand, deep learning-based ROMs leverage neural networks to account for nonlinear dynamics. It's crucial to differentiate between classical ROM approaches and those utilizing deep learning. The former uses linear algebra and statistical techniques to find a reduced subspace that encapsulates the most significant dynamics of the system. Nevertheless, deep learning-based ROMs are part of broader field of representation learning [23, 24, 25, 26, 27], specifically state representation learning (SRL) [28, 29], where high dimensional states are compressed into lower dimensional representations that evolves with time.

### 2.2 CO2 Sequestration Optimization

While existing literature has different focus in terms of optimizing CO2 storage, e.g. maximizing CO2 storage capacity and efficiency [30, 31], minimize potential geomechanical risks that can lead to CO2 leakage [32], or a combination of these [33, 34]. The objective of CO2 sequestration in this research focuses on associated financial gains, which is given by

$$NPV(u) = \sum_{t=1}^{N_t} [\gamma(\Delta t)]^t * (\sum_{j=1}^{N_{inj}} (r_{credit} - r_{opr}) q_t^{inj,j} + \sum_{i=1}^{N_{prod}} (-r_w) q_{t,w}^i + \sum_{i=1}^{N_{prod}} (-r_{co2}) q_{t,g}^i) * \Delta t$$

(5)

Here $u$ is the well control sequences ($u_1, u_2, ..., u_{N_t}$), where each control variable entails the CO2 injection rates for each injection well and bottom-hole pressure for every production well. $N_t$ denotes the timesteps of the simulation, $\gamma$ is the depreciation factor every $\Delta t$ days, $N_{inj}$ and $N_{prod}$ are number of injection wells and production wells, respectively; $r_{credit}$ in USD/ton is the carbon tax credit for every metric ton CO2 stored, $r_{opr}, r_w, r_{co2}$ are the operational costs of CO2 injection (USD/ton), brine handling cost (USD/STB) and CO2 handling cost (USD/ton), respectively; $q_t^{inj,j}$ in unit of ton/day is the CO2 injection rate of injection well $j$, $q_{t,w}^i$ in unit of STB/day is the brine production rate, and $q_{t,g}^i$ in unit of ton/day is the CO2 production rate of each production well $i$. Note that $q_t^{inj,j}$ is part of control variable, while $q_{t,w}^i$ and $q_{t,g}^i$ are system outputs, which is dependent on control variables.

The optimization problem can be formulated as

$$\underset{u}{\arg\max}\, NPV(u)$$

(6)

Subject to

$$x_{t+1} = f(x_t, u_t), y_{t+1} = g(x_{t+1}, u_t)$$

In addition, control variables are subject to engineering constraints $u_{low} \leq u \leq u_{up}$, $u_{low}$ and $u_{up}$ denotes the lower and upper bound of control variable.

## 3    Methodology

### 3.1    Embed to control and observe (E2CO)

The original Embed-to-Control (E2C) is a variational autoencoder-based representation learning model for stochastic optimal control [13]. It has been modified to a deterministic form for surrogate modeling in reservoir simulation [11]. This model consists of an encoder network that projects high-dimensional state spaces (saturation, pressure, etc.) into a lower latent space, a transition model that governs the evolution of this latent space from current time point to the next, and a decoder that reconstructs states of next time from latent representation of corresponding time step. Advancing from E2C, an enhanced architecture named Embed to Control and Observe (E2CO) has been introduced [12], which incorporates the well observation formulation directly within the latent space. Comparative studies suggest E2CO's superiority over E2C in predicting well data and state estimations, largely attributable to E2CO's direct latent space regularization via well observation data. The graphical model of E2CO is depicted in Figure 1.

From a mathematical standpoint, the encoder-transition-decoder architecture, supplemented by an output network which calculates well observations, is detailed as follows:

Encoder: $\qquad\qquad\qquad\qquad z_t = Q_\varphi(x_t)$

Transition: $$\hat{z}_{t+1} = F_{\omega_1}(z_t, u_t)$$

Output: $$\hat{y}_{t+1} = G_{\omega_2}(\hat{z}_{t+1}, u_t)$$

Decoder: $$\hat{x}_{t+1} = P_\theta(\hat{z}_{t+1})$$

In E2C and E2CO, in order to employ locally linear relation in the latent space, the transition network is partitioned to two sub-networks to compute local matrices $A(z_t)$ and $B(z_t)$, which is expressed as

$$A(z_t) = \omega_A h^{trans}_{\omega_{1,1}}(z_t) + b_A$$
$$B(z_t) = \omega_B h^{trans}_{\omega_{1,1}}(z_t) + b_B$$

(7)

The next latent state can be computed as $\hat{z}_{t+1} = A(z_t) * z_t + B(z_t) * u_t$. This approach signifies that the transition network's trainable weights are a composition of weights from these operations, suggesting $\omega_1 = (\omega_{1,1}, \omega_A, \omega_B, b_A, b_B)$.

Distinct from E2C, E2CO incorporates system output formulation in the latent space. For the output network in E2CO, local matrices $C(z_t)$ and $D(z_t)$ are estimated directly using the latent states, given by

$$C(z_t) = \omega_C h^{trans}_{\omega_{2,1}}(z_t) + b_C$$
$$D(z_t) = \omega_D h^{trans}_{\omega_{2,1}}(z_t) + b_D$$

(8)

And the system outputs are calculated by $\hat{y}_{t+1} = C(z_t) * \hat{z}_{t+1} + D(z_t) * u_t$. Likewise, $\omega_2 = (\omega_{2,1}, \omega_C, \omega_D, b_C, b_D)$.

The whole network is trained by minimizing the loss defined as follows,

$$\mathcal{L}_{total,i} = \lambda_{rec}\mathcal{L}_{rec,i} + \lambda_{kl}\mathcal{L}_{KL,i} + \lambda_{yobs}\mathcal{L}_{yobss,i}$$

(9)

Where $\mathcal{L}_{total,i}, \mathcal{L}_{rec,i}, \mathcal{L}_{KL,i}, \mathcal{L}_{yobss,i}$ denotes the total loss, reconstruction loss, KL-divergence loss, and observation data loss for each sample $i$, respectively. $\lambda_{rec}, \lambda_{kl}$, and $\lambda_{yobs}$ are associated weights used to balance those loss. Here $\mathcal{L}_{rec,i} = \|x_{t,i} - \hat{x}_{t,i}\|_2 + \|x_{t+1,i} - \hat{x}_{t+1,i}\|_2$, $\mathcal{L}_{KL,i} = \|z_{t,i}\|_2 + \|z_{t+1,i} - \hat{z}_{t+1,i}\|_2$ and $\mathcal{L}_{yobss,i} = \|y_{t+1,i} - \hat{y}_{t+1,i}\|_2$. While incorporating physics loss has been suggested to improve sample efficiency, this study does not explore that avenue since a comprehensive knowledge of underlying physics is required, as discussed in our previous studies [35, 36].

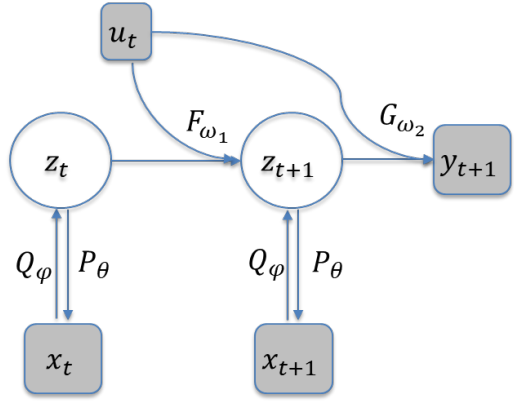

Figure 1. Graphical model of E2CO, gray boxes represent the training tuple $\{x_t, u_t, x_{t+1}, y_{t+1}\}$. $Q_\varphi$ is the encoder network parameterized by $\varphi$, while $P_\theta$ signifies a decoder network parameterized by $\theta$. The transition network, expressed as $F_{\omega_1}$, maps from $z_t$ to $z_{t+1}$ given $u_t$. In actual implementation, it is further parameterized using two networks to approximate $A$ and $B$ so that the locally linear equation can be used. Likewise, $G_{\omega_2}$ denotes the output network designed to approximate $C$ and D. It has been observed that linear transition structures tend to achieve faster convergence than their nonlinear counterparts during the training phase.

### 3.2  Reinforcement learning (RL)

Reinforcement learning has found applications across a broad range of fields, to name a few, robotics control and trajectory optimization [37, 38], autonomous driving [39, 40], large language models [41] and so forth. Attempts of employing RL in optimizing developing subsurface resources have also been explored [42, 43, 44, 45]. The objective of RL is to learn an optimal policy $\pi^*$ for the RL agent that maximize the cumulative reward collected by repeatedly interacting with the environment, which can be expressed as:

$$\pi^* = \underset{\pi}{\mathrm{argmax}}\, Q_\pi(s, a)$$

*(10)*

where $\pi$ is the policy. It is a mapping from states to actions. In other words, policy determines the agent's behavior from a specific state. $Q_\pi(s, a)$ refers to the action-value function, or Q function when executing policy $\pi$. The RL agent makes action $a$ according to the policy under certain state $s$, which read $a \sim \pi(a|s)$. The action-value function $Q_\pi(s, a)$ describes the expected return of taking action $a$ in a state $s$ and following a specific policy $\pi$ thereafter. Mathematically,

$$Q_\pi(s, a) = \mathbb{E}_\pi[R_t | s_t = s, a_t = a]$$

*(11)*

where $R_t = \sum_{k=0}^{\infty} \gamma^k r_{t+k+1}$ is the cumulative discounted reward starting from state s, taking action a, and complying with policy $\pi$.

To use reinforcement learning for optimization of CO2 sequestration, designing a suitable reward function is needed. To this end, the immediate reward at time $t$ is defined as the financial gain at that time:

$$r_t = \sum_{j=1}^{N_{inj}} (r_{credit} - r_{opr}) q_t^{inj,j} + \sum_{i=1}^{N_{prod}} (-r_w) q_{t,w}^i + \sum_{i=1}^{N_{prod}} (-r_{co2}) q_{t,g}^i$$

(12)

Note that when training the RL agent, scaling of $r_t$ will not impact the results. We show the general framework of RL in Figure 2.

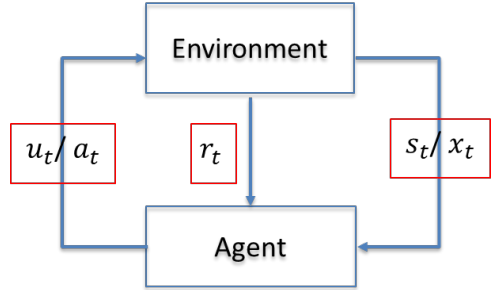

Figure 2. Sketch plot of general reinforcement learning (RL) framework. A RL agent determines actions $a_t$ or $u_t$, based on the current state $s_t$ or $x_t$, in order to maximize cumulative rewards $\sum_t^\infty r_t$.

The Deep Q-Network (DQN) [46] emerges in the literature as a prominent model-free reinforcement learning technique suited for tasks within discrete action spaces. DQN implements the policy by selecting the action associated with the highest Q-value. The method benefits from the ability to learn from past experiences, however, DQN's ability is limited due to challenges in adapting it for continuous or high-dimensional action spaces. Conversely, policy-based approaches, e.g. policy gradient method [47] and Asynchronous Advantage Actor-Critic (A3C) [48] are adept at handling continuous action spaces. These methods directly parameterize the policy, which can be either stochastic or deterministic. However, because these methods update the policy throughout the learning process, reutilizing previous experience data becomes challenging, particularly in on-policy methods. To enhance data efficiency, Proximal Policy Optimization (PPO) [49] was introduced, which uses past data samples to update policy parameters by imposing penalties for significant deviations from previous policies. Despite its advancements, PPO remains an on-policy method. A promising avenue for improving data efficiency is actor-critic methods, which integrate aspects of both value-based and policy-based approaches. A notable breakthrough in continuous control problems is the Deep Deterministic Policy Gradient (DDPG) [50], which excels in utilizing off-policy updates with replay data and optimizing the policy parameters based on the Q-value function gradient. Nonetheless, DDPG struggles with exploration and is highly sensitive to hyperparameter settings, often requiring additional adjustments for optimal performance.

Soft Actor-Critic (SAC) [51] emerged as a solution to the exploration-exploitation challenge by incorporating the entropy of the policy into the reward function, encouraging the agent to explore more effectively without sacrificing performance. SAC is an off-policy algorithm that combines the benefits of actor-critic methods with those of maximizing entropy—a measure of randomness in the policy—to achieve robust and efficient learning, especially in complex environments with high-dimensional, continuous action spaces. SAC employs an actor-critic architecture with two Q-functions (critics), which are represented by two neural networks parameterized by $\phi_1$ and $\phi_2$ respectively, to minimize overestimation bias, and a policy network (actor) which parameterized by $\chi$ that outputs actions from a set of available actions. The Q functions are updated to minimize the following loss functions:

$$J_{Q,i} = \mathbb{E}_{(s,a,r,s')\sim\mathcal{D}}[(Q_{\phi_i}(s,a) - (r + \gamma(\min_{j=1,2} Q_{\phi_{targ,j}}(s',a') - \alpha\log\pi_\chi(a'|s'))))^2]$$

(13)

Where $\mathcal{D}$ refers to replay buffer, $Q_{\phi_i}(s,a)$ is the neural network output, r refers to reward, $\gamma$ is the discount factor as defined previously. $a'\sim\pi_\chi(\cdot|s')$ represents next action under policy $\pi$. The last term signifies the entropy of policy $\pi$, and $\alpha$ is called the temperature parameter that determines the importance of entropy term. For a deterministic policy, the entropy is 0.

The policy parameters are updated by maximizing the expected return and entropy, which reads,

$$J_\pi = \mathbb{E}_{s\sim\mathcal{D}, a\sim\pi_\chi}[\min_{i=1,2} Q_{\phi_i}(s,a) - \alpha\log\pi_\chi(a|s)]$$

(14)

The algorithm of SAC is presented in Algorithm 2 in section 3.5.

### 3.3 Surrogate model-based reinforcement learning

In the context of practical CO2 storage operations, state spaces characterizing pressure or CO2 mole fraction may be of extremely high dimension. Using reinforcement learning (RL) in high-dimensional state spaces presents challenges that can hinder learning efficiency and effectiveness. Instead of dealing with high-dimensional states, a lower-dimensional latent space simplifies the RL problem, as the agent has fewer dimensions to explore. To this end, we introduce the surrogate model-based RL that is shown in Figure 3. With high-dimensional states $s_t$ or $x_t$ being substituted by $z_t$, the system outputs in the reward function are also replaced by those forecasted through trained E2CO model. Consequently, the reward function $\hat{r}_t$ becomes:

$$\hat{r}_t = \sum_{j=1}^{N_{inj}}(r_{credit} - r_{opr})q_t^{inj,j} + \sum_{i=1}^{N_{prod}}(-r_w)\hat{q}_{t,w}^i + \sum_{i=1}^{N_{prod}}(-r_{co2})\hat{q}_{t,g}^i$$

(15)

Note that $\hat{q}_{t,w}^i$ and $\hat{q}_{t,g}^i$ are proxy well observations in $\hat{y}_t = \begin{bmatrix} \hat{q}_{t,w} \\ \hat{q}_{t,g} \\ \hat{P}_{wf,t} \end{bmatrix} \in R^{2*N_{prod}+N_{inj}}$ from trained E2CO model.

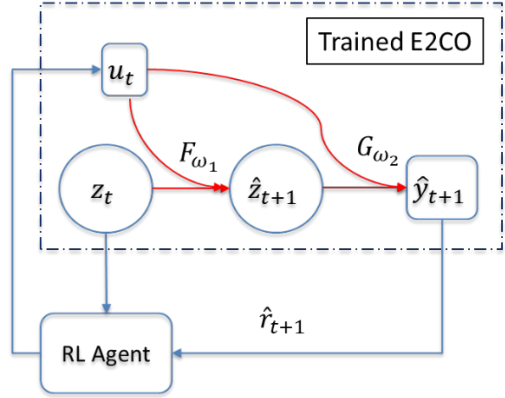

Figure 3. E2CO-based reinforcement learning at time $t$. The dashed box indicates the trained E2CO model. The RL agent receives $z_t$ and output control action $u_t$ by interacting with E2CO. The action $u_t$ propels the environment to a new latent state $\hat{z}_{t+1}$, obtaining approximate well observation quantities $\hat{y}_{t+1}$ and consequently reward $\hat{r}_{t+1}$. This process happens recursively until final $N_t$ is reached.

### 3.4   Data acquisition and preprocessing

We collect data used to train and test the surrogate model in a similar manner as in [11] and [12]. However, it's essential to underscore the distinctions in our model which, in turn, influenced the design of our reinforcement learning (RL) framework. We utilized a commercial compositional reservoir simulator CMG GEM [52] to model and simulate the CO2 injection scenarios. The proposed framework is implemented within a compositional simulation model where CO2 is injected into a saline aquifer. The model consists of $64 \times 64 \times 1$ grid blocks, each of which has dimension of 65.6 ft $\times$ 65.6 ft $\times$ 65.6 ft in x, y and z directions. We simulated scenarios where we inject CO2 for permanent storage using 4 injector wells, while 5 producer wells are used for pressure maintenance. This model is run for 2000 days and well controls including injection rates and bottom hole pressure (BHP) vary every 100 days. The permeability field and the well locations are plotted in Figure 4.

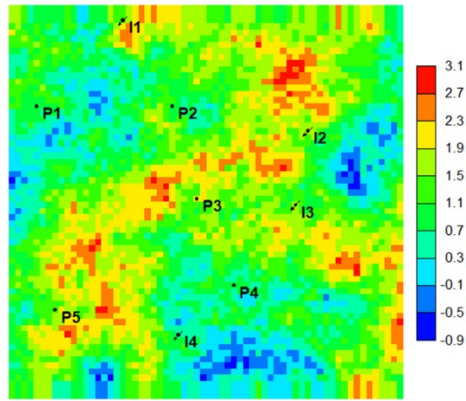

Figure 4. Log permeability map, in mD; P1-P5 indicates 5 producers and I1-I4 refers to 4 injectors.

Other important physical properties used in the simulation are listed in Table 1.

Table 1. Other properties used in simulation setup

| Porosity | 0.13 | Well radius, ft | 0.30 |
|---|---|---|---|
| Reservoir Depth, fts | 7500 | skin factor | 0.0 |
| Reservoir Temperature, F | 200.0 | Rock compressibility, $psia^{-1}$ | $4.0 \times 10^6$ |

Each simulation run is referred as one trajectory, or one episode. In this paper, we simulate a total of 600 trajectories by randomly varying the well controls. Considering engineering constraints, we set BHP limits between 2200 and 2500 psi and injection rates ranging from $10^5$ to $10^6$ cubic feet under standard surface condition (temperature 60 F and atmosphere pressure). Starting from identical initial states for all simulations, we collect 20 data pairs per trajectory, culminating in 12,000 data pairs overall $D_{total} = \{(x_0, u_0, x_1, y_1), (x_1, u_1, x_2, y_2), ...\}$. This dataset is then divided into training and testing dataset in a 3:1 ratio.

Before training the surrogate model, the data undergo normalization to ensure all quantities are scaled to fall within the range [0, 1]. The normalization formula is given by

$$p_{norm} = \frac{p - p_{min}}{p_{max} - p_{min}}$$

$$q_{w,norm} = \frac{q_w - q_{w,min}}{q_{w,max} - q_{w,min}}$$

$$q_{g,norm} = \frac{q_g - q_{g,min}}{q_{g,max} - q_{g,min}}$$

*(16)*

Where $p$ denotes all pressure quantities in $x_t$, $u_t$ and $y_t$. $q_w$ is the water rate which is in $y_t$. $q_g$ is the gas rate in both $u_t$ and $y_t$. $p_{min}, p_{max}, q_{w,min}, q_{w,max}, q_{g,min}, q_{g,max}$ typically represent the minimum and maximum values observed in the dataset for corresponding parameters, although this is not an absolute requirement. This scaling is crucial for maintaining consistency across different variables and facilitating the model's learning process.

Accordingly, during the reward calculation within the RL framework, it's essential to revert all normalized quantities back to their original values. Moreover, to ensure actions chosen by the RL policy network remain within feasible limits, incorporating a scaling layer that adjusts the action values back to permissible ranges is crucial. This ensures that the actions are realistic and applicable to the actual operational context.

### 3.5 Workflow

The workflow of this work is picturized in Figure 5. The workflow of this framework can be summarized as follows: Firstly, we perturb the well controls to obtain states and well observations at various timesteps, these states and observations then be assembled to training tuple $(x_t, u_t, x_{t+1}, y_{t+1})$ to train the following E2CO network. Note that the $x_t$ entails pressure and CO2 mole fraction at current step, while $x_{t+1}$ comprises pressure and CO2 mole fraction at next time step after applying well control $u_t$. Subsequently, this E2CO network, once trained, serves as a proxy environment for the reinforcement learning (RL) agent to interact with. The RL agent's role is to identify the most financially beneficial well control strategies.

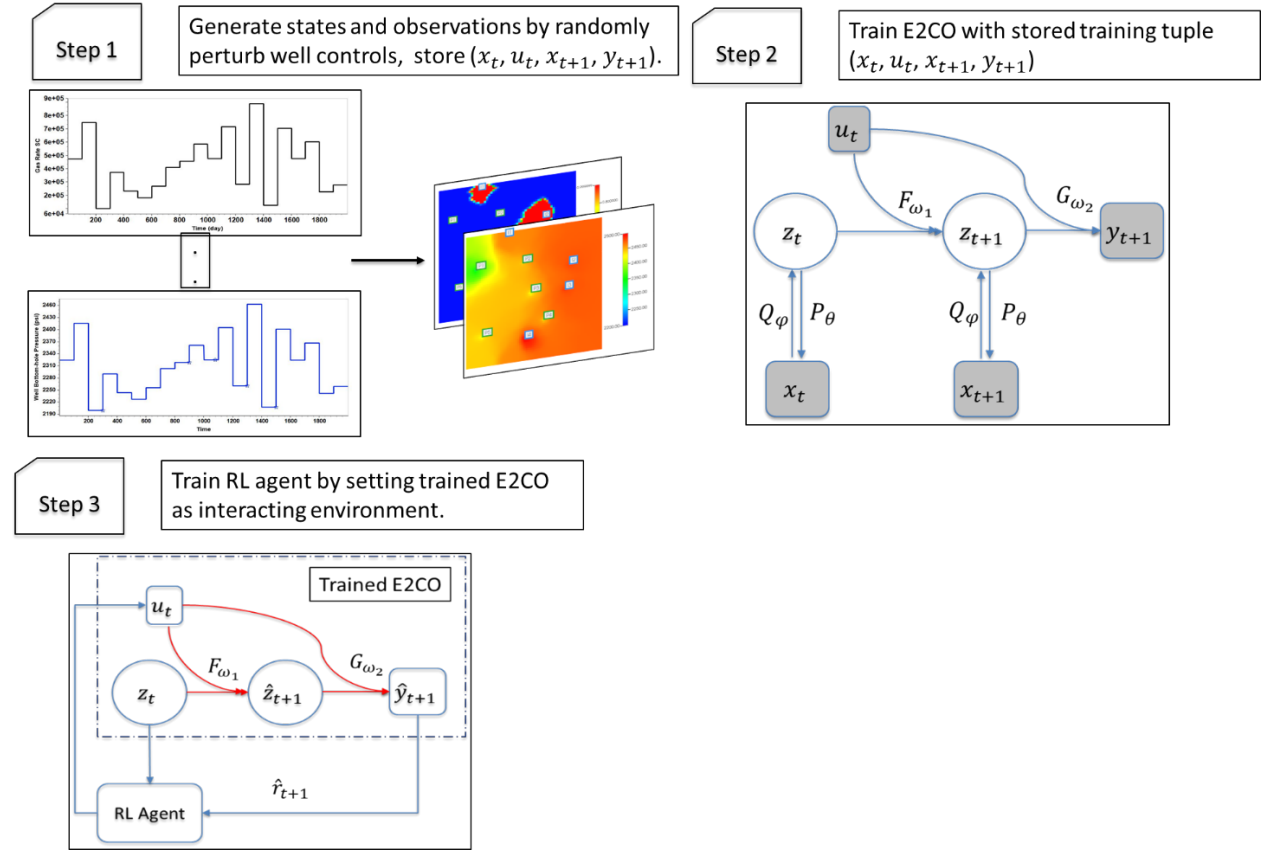

Figure 5. Workflow of proposed approach.

We train the E2CO model according to algorithm 1, following that we use the trained E2CO model as interacting environment for the RL agent, the RL agent is optimized for maximum NPV using the Soft Actor-Critic algorithm, as shown in algorithm 2.

**ALGORITHM 1: TRAINING E2CO PROXY MODEL**

**Input**: training tuples $\{x_t, u_t, x_{t+1}, y_{t+1}\}$, initial neural network weights $\Theta = \{\varphi, \omega_1, \omega_2, \theta\}$ including encoder $Q_\varphi$, transition network $F_{\omega_1}$, output network $G_{\omega_2}$, and decoder $P_\theta$, learning rate $\eta$, and total training epochs $N_{epoch}$

**for** k=1 to $N_{epoch}$ **do**

$z_t = Q_\varphi(x_t), \hat{z}_{t+1} = F_{\omega_1}(z_t, u_t), \hat{x}_{t+1} = P_\theta(\hat{z}_{t+1}), \hat{y}_{t+1} = G_{\omega_2}(\hat{z}_{t+1}, u_t)$ and $z_{t+1} = Q_\varphi(x_{t+1})$

Calculate $\mathcal{L}_{total}$ according to defined formulation in section 3.1.

$\nabla\Theta_k \leftarrow Backprop(\mathcal{L}_{total})$

$\Theta_k \leftarrow optimizer.step()$       # update network weights using predefined optimizer

**end for**

**Output**: optimized model parameters $\Theta$

| | |
|---|---|
| **ALGORITHM 2: OPTIMIZATION WITH SOFT ACTOR CRITIC (SAC) ALGORITHM** | |
| **Input**: | trained E2CO model $\Theta = \{\varphi, \omega_1, \omega_2, \theta\}$ including encoder $Q_\varphi$, transition network $F_{\omega_1}$, output network $G_{\omega_2}$, and decoder $P_\theta$, set environment as $env \leftarrow E2CO_\Theta$. Policy network (actor) parameterized by $\chi$; two Q-value network (critics) parameterized by $\phi_1$ and $\phi_2$, and their target networks $\phi_{targ,1}$ and $\phi_{targ,2}$. |
| **Procedure**: | Initialize policy network, Q-value networks, and target Q-value networks. <br> Compute initial latent state $z_0 = Q_\varphi(x_0)$ <br> **for** each iteration **do** <br>     **for** each environment step **do** <br>         Sample $u_t$ from policy network given current latent state $z_t$, $u_t \sim \pi_\chi(\cdot \| z_t)$ <br>         Execute $u_t$ in the environment $env$, $z_{t+1}, r_t = env(z_t, u_t)$ <br>         Store $(z_t, u_t, r_t, z_{t+1})$ in a replay buffer $\mathcal{D}$ <br>     **end for** <br>     **for** each gradient step **do** <br>         Sample mini-batch of $\{(z_t, u_t, r_t, z_{t+1}), \ldots\}$ from the replay buffer <br>         $\phi_i \leftarrow Backprop(J_{Q,i})$      Update critics by minimizing the loss $J_{Q,i}$ for $i = 1,2$ <br>         $\chi \leftarrow Backprop(J_\pi)$      Update actor by minimizing the loss $J_\pi$ <br>         $\phi_{targ,i} \leftarrow \tau\phi_i + (1-\tau)\phi_{targ,i}$ Update target Q-value networks for $i = 1,2$ <br>     **end for** <br> **end for** |
| **Output**: | optimized parameters $\phi_1$, $\phi_2$ and $\chi$ |

## 4     Results

In the following subsections, we firstly show the performance of E2CO for CO2 storage operations, then we present an optimization case where injection rates and bottom hole pressure has been optimized for CO2 storage.

### 4.1.    E2CO for CO2 sequestration

Figure 6 and 7 compares the overall CO2 mole fraction $z_{CO_2}$ and fluid pressure $p$ derived from E2CO model against those obtained from high-fidelity simulations. The top row of Figure 6 shows $\hat{z}_{CO_2}$ predictions of one testing control strategies given initial $z_{CO_2,0}$ using E2CO. From left to right, 10 snapshots at different times are presented. The middle row of Figure 6 depicts the evolution of $z_{CO_2}$ for the same control strategies and initial conditions using CMG IMEX. The bottom row displays the disparities between the forecasted $\hat{z}_{CO_2}$ and reference $z_{CO_2}$ at chosen time. Similarly, in Figure 7, we present the predicted fluid pressure $\hat{p}$, reference fluid pressure $p$ and their difference at the top, middle and bottom rows, respectively.

As can be observed from Figure 6 and 7, the E2CO model achieves satisfactory performance when predicting the evolution of $z_{CO_2}$ as well as fluid pressure $p$. It is noted that the error in mole fraction incrementally grow over time from left to right, which is attributable to the sequential prediction methodology employed in this framework. Regarding the pressure predictions, it's showing decent performance, with around 1 % relative error in the pixel level. We emphasize that besides the computational efficiency the surrogate model brings to the RL optimization framework, it also facilitates the monitoring of CO2 plume movement and the potential over-pressurization during CO2 storage operations.

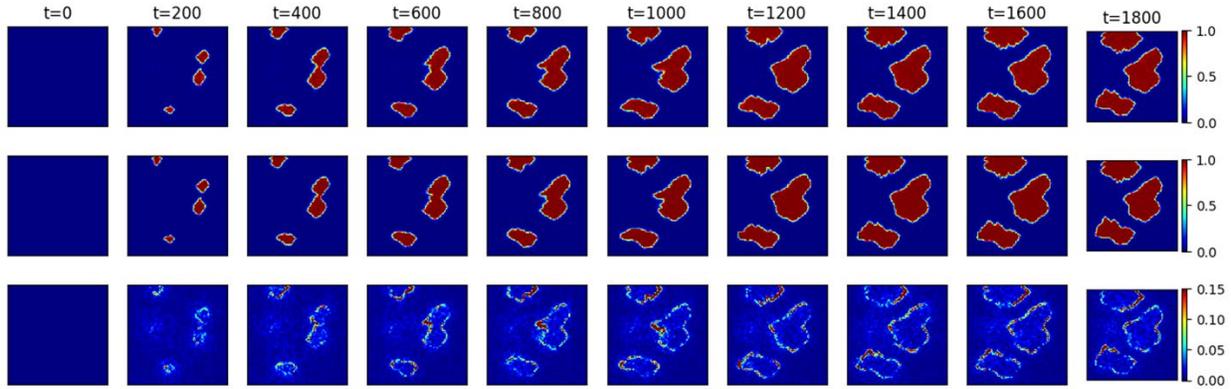

Figure 6. (top row): predicted $\hat{z}_{CO_2}$ by E2CO at selected days; (middle row): ground truth $z_{CO_2}$ at corresponding days; (bottom row): absolute error between prediction and reference $|\hat{z}_{CO_2} - z_{CO_2}|$.

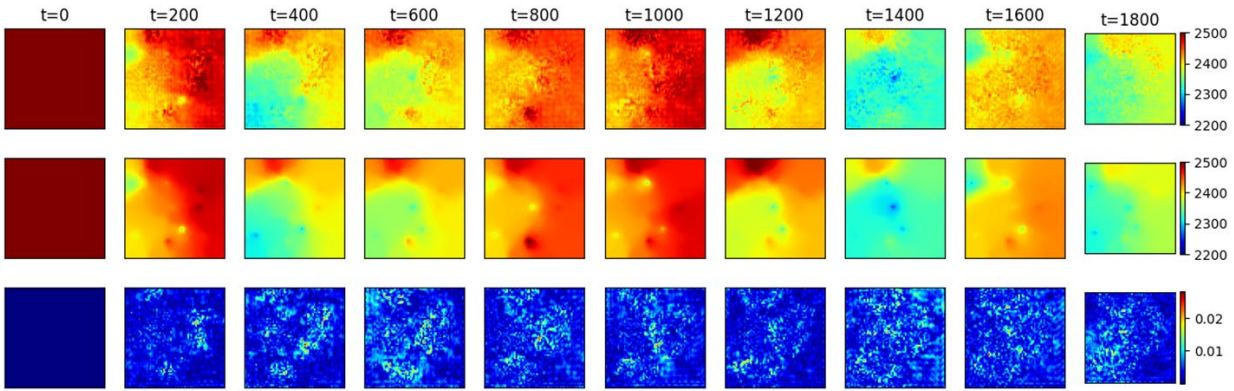

Figure 7. (top row): predicted fluid pressure $\hat{p}$ in unit of psia by E2CO at selected days; (middle row): ground truth $p$ in unit of psia at corresponding days; (bottom row): absolute relative error between the prediction and reference $\frac{|\hat{p}-p|}{p}$.

Figure 8 showcases the comparison of the well observation quantities over time predicted with E2CO model for 5 extractors against those from high-fidelity simulations. The right column of Figure 8 displays water rates from 5 production wells, while the left column shows the gas rate (indicator of gas leakage) at corresponding production wells. Within the plots, blue curves represent the predictions with E2CO model, while orange curves refer to the reference quantities from simulators. It can be observed that they exhibit a high degree of concordance. Additionally, it is noted that in the absence of optimization measures, production wells 3 and 5 are particularly prone to significant gas leakage.

Table 2 showcases the computation time of full-order compositional simulation, the training of the surrogate model and inference using trained surrogate model. Each full-order simulation costs 120 to 130 seconds on a CPU with an Intel(R) Xeon(R) W-2223 CPU @3.60GHz. The training takes around 3 hours (10800 seconds) when the training epoch is 200 and batch size is 4. Inference is 2.86 seconds for 100 testing samples. Both training and inference are run on a NVIDIA A100 provided by Texas A&M High-Performance Research Computing.

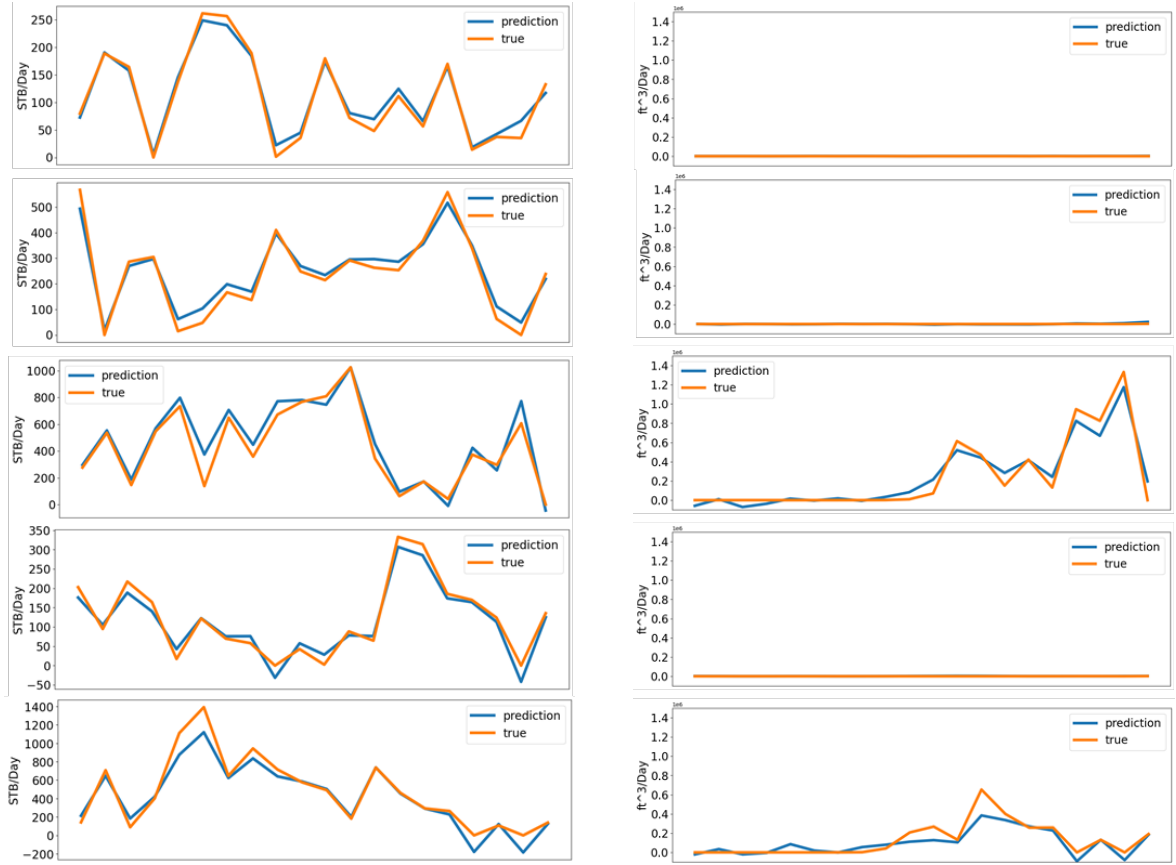

Figure 8. (Left) Water production ($STD/day$) from 5 producers, (Right) Gas production ($ft^3/day$) from 5 producers, with the y axis showing scale of $10^6$. Blue curve refers to predictions from E2CO, while orange curve represents the ground truth from simulation.

Table 2. Comparison of computational time

| **Full-order Simulations** | **E2CO** | |
|---|---|---|
| | Training time | Inference |
| 12000~13000 sec (for 100 simulation runs, using CPU) | ~ 3 hours (200 epochs, batch size 4, using GPU A100) | 2.86 sec (for 100 testing samples, using GPU A100) |

## 4.2 Optimization of CO2 injection into saline aquifer

In this case, we assume $r_{credit} = 50$ \$/ton, $r_{opr} = 10$ \$/ton, $r_w = 5$ \$/STB and $r_{co2} = 50$ \$/ton. The discount factor $\gamma$ is set to be 0.986 for $\Delta t = 100$ days. Figure 9 displays the optimized bottom hole pressures (BHPs) for production wells alongside the optimized daily injection rates for CO2 injectors. It's noticeable that the production wells tend to maintain high BHPs, aiming to prevent gas breakthrough. On the other hand, except for injector well 2, which starts with a high injection volume and then shifts to more conservative rates after day 600, all injector wells maintain relatively low injection rates. While for the injectors, all injection wells except well 2 are keeping a low injection rate. Injector 2 is operating at high injection volume at the beginning, followed by conservative injection rates after day 600. Injection wells numbered 1 and 3 exhibit consistently low injection rates during the storage period, while well 4 shows a minor increase between days 900 and 1100.

Figure 10 displays the CO2 plume before and after implementing optimization procedures. Prior to optimization, wells 3 and 5 experienced significant gas breakthrough, with wells 2 and 4 also at high risk of leakage. Post-optimization, all wells were safeguarded against gas breakthrough, except for well 3, which is on the brink of gas production, likely due to an equilibrium sought between maximizing CO2 storage and avoiding CO2 leakage in that area.

Table 3 presents the quantitative results of optimal NPV at the end of storage period with various frameworks. Utilizing the reduced-order model-based reinforcement learning (ROM-RL) approach yields a final NPV of \$27.7 million, in contrast to a -\$10.3 million NPV when employing full-order models. This disparity is attributed to differences in well observation quantities by these two models. Regardless, both approaches significantly outperform the baseline scenario, where well controls are set arbitrarily, demonstrating their effectiveness in rendering the CO2 storage project economically viable.

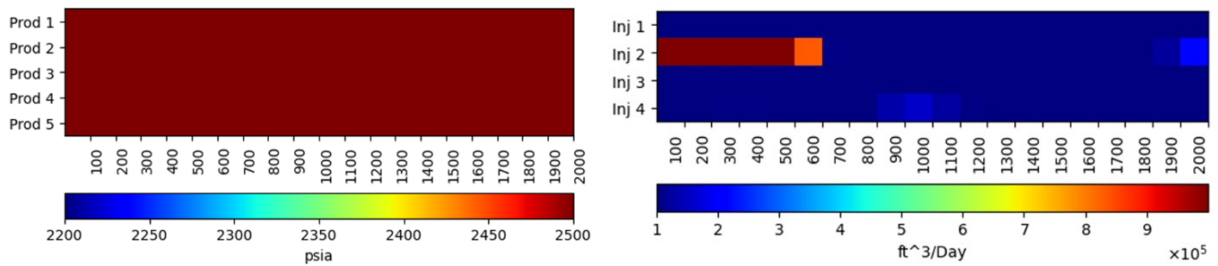

Figure 9. Optimal well control strategies for 2000 days. Left is the optimal BHPs for 5 producers. Right is the optimal injection rates for 4 injectors.

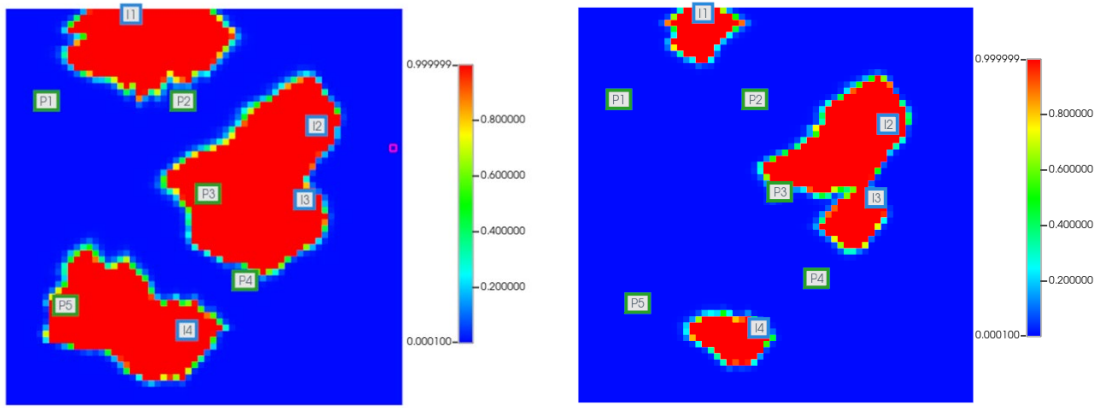

Figure 10. CO2 plume before (left) and after (right) optimized well control strategies.

Table 3. NPV comparison at the end of CO2 storage period

| Optimal NPV (ROM-RL) | Optimal NPV (Full-order) | NPV (base case) |
|---|---|---|
| $27.7m | ($10.3m) | ($4329.4 m) |
| parentheses () indicates negative amount. | | |

## 5    Concluding remarks

In this work, we proposed a novel surrogate model-based reinforcement learning framework for optimizing pressure management strategies in CO2 sequestration, aiming to maximize financial returns for storage projects. Surrogate model-based reinforcement learning aims to learn a proxy model to mimic the environment dynamics in the highly compressed representation space. These models are simpler and less computationally demanding than 'true' models but still capture enough of the environment's dynamics to allow for effective planning and decision-making. This proposed approach shows efficient learning by combining the sample efficiency of model-based methods with the simplicity and computational tractability of model-free methods.

The proposed framework was tested on a case where CO2 is injected into saline aquifer. Without performing optimization measures, a few production wells are undergoing severe gas leakage, leading to a substantial loss of $4329.4 million in NPV. However, utilizing the surrogate model-based RL framework for optimization yielded a positive NPV of $27.7 million. When the optimized controls were applied in the full-order compositional simulator, an NPV of -$10.3 million is obtained, which further proves that the surrogate model translates well to the real world model.

The framework's versatility allows for the optimization of diverse control types, including both categorical and discrete variables. It supports heuristic approaches, e.g. when the pressure at certain region exceed the threshold, we can impose a large penalty in the reward function to avoid the fault reactivation or caprock fracturing. In future work, this framework can be extended to co-optimizing well placement, well controls and others.

**ACKNOWLEDGEMENTS**

**DATA AVAILABILITY STATEMENTS**

The datasets and code will be available at https://github.com/jungangc/CCS_E2CO-RL once published.

**DECLARATIONS**

The authors have no relevant financial or non-financial interests to disclose.

**REFERENCES**

[1] Jha, B., & Juanes, R. (2014). Coupled multiphase flow and poromechanics: A computational model of pore pressure effects on fault slip and earthquake triggering. *Water Resources Research*, *50*(5), 3776-3808.

[2] Cappa, F., & Rutqvist, J. (2011). Modeling of coupled deformation and permeability evolution during fault reactivation induced by deep underground injection of CO2. *International Journal of Greenhouse Gas Control*, *5*(2), 336-346.

[3] González-Nicolás, A., Cihan, A., Petrusak, R., Zhou, Q., Trautz, R., Riestenberg, D., ... & Birkholzer, J. T. (2019). Pressure management via brine extraction in geological CO2 storage: Adaptive optimization strategies under poorly characterized reservoir conditions. *International Journal of Greenhouse Gas Control*, *83*, 176-185.

[4] Kempka, T., Nielsen, C. M., Frykman, P., Shi, J. Q., Bacci, G., & Dalhoff, F. (2015). Coupled hydro-mechanical simulations of CO2 storage supported by pressure management demonstrate synergy benefits from simultaneous formation fluid extraction. *Oil & Gas Science and Technology–Revue d'IFP Energies nouvelles*, *70*(4), 599-613.

[5] Jansen, J. D., & Durlofsky, L. J. (2017). Use of reduced-order models in well control optimization. Optimization and Engineering, 18, 105-132.

[6] Van Doren, J. F., Markovinović, R., & Jansen, J. D. (2006). Reduced-order optimal control of water flooding using proper orthogonal decomposition. *Computational Geosciences*, *10*, 137-158.

[7] Jin, Z. L., & Durlofsky, L. J. (2018). Reduced-order modeling of CO2 storage operations. *International Journal of Greenhouse Gas Control*, *68*, 49-67.

[8] Pawar, S., Rahman, S. M., Vaddireddy, H., San, O., Rasheed, A., & Vedula, P. (2019). A deep learning enabler for nonintrusive reduced order modeling of fluid flows. *Physics of Fluids*, *31*(8).

[9] Pant, P., Doshi, R., Bahl, P., & Barati Farimani, A. (2021). Deep learning for reduced order modelling and efficient temporal evolution of fluid simulations. *Physics of Fluids*, *33*(10).

[10] Fresca, S., Dede', L., & Manzoni, A. (2021). A comprehensive deep learning-based approach to reduced order modeling of nonlinear time-dependent parametrized PDEs. *Journal of Scientific Computing*, *87*, 1-36.

[11] Jin, Z. L., Liu, Y., & Durlofsky, L. J. (2020). Deep-learning-based surrogate model for reservoir simulation with time-varying well controls. *Journal of Petroleum Science and Engineering*, *192*, 107273.

[12] Coutinho, E. J. R., Dall'Aqua, M., & Gildin, E. (2021). Physics-aware deep-learning-based proxy reservoir simulation model equipped with state and well output prediction. *Frontiers in Applied Mathematics and Statistics*, *7*, 651178.

[13] Watter, M., Springenberg, J., Boedecker, J., & Riedmiller, M. (2015). Embed to control: A locally linear latent dynamics model for control from raw images. *Advances in neural information processing systems*, *28*.

[14] Zhang, M., Vikram, S., Smith, L., Abbeel, P., Johnson, M., & Levine, S. (2019, May). Solar: Deep structured representations for model-based reinforcement learning. In *International conference on machine learning* (pp. 7444-7453). PMLR.

[15] Levine, N., Chow, Y., Shu, R., Li, A., Ghavamzadeh, M., & Bui, H. (2019). Prediction, consistency, curvature: Representation learning for locally-linear control. *arXiv preprint arXiv:1909.01506*.

[16] Shu, R., Nguyen, T., Chow, Y., Pham, T., Than, K., Ghavamzadeh, M., ... & Bui, H. (2020, November). Predictive coding for locally-linear control. In *International Conference on Machine Learning* (pp. 8862-8871). PMLR.

[17] He, J., & Durlofsky, L. J. (2014). Reduced-order modeling for compositional simulation by use of trajectory piecewise linearization. SPE Journal, 19(05), 858-872.

[18] Voskov, D. V., & Tchelepi, H. A. (2012). Comparison of nonlinear formulations for two-phase multi-component EoS based simulation. Journal of Petroleum Science and Engineering, 82, 101-111.

[19] Tan, X., Gildin, E., Florez, H., Trehan, S., Yang, Y., & Hoda, N. (2019). Trajectory-based DEIM (TDEIM) model reduction applied to reservoir simulation. *Computational Geosciences*, *23*, 35-53.

[20] Trehan, S., & Durlofsky, L. J. (2016). Trajectory piecewise quadratic reduced-order model for subsurface flow, with application to PDE-constrained optimization. *Journal of Computational Physics*, *326*, 446-473.

[21] Rathinam, M., & Petzold, L. R. (2003). A new look at proper orthogonal decomposition. *SIAM Journal on Numerical Analysis*, *41*(5), 1893-1925.

[22] Ravindran, S. S. (2000). A reduced-order approach for optimal control of fluids using proper orthogonal decomposition. *International journal for numerical methods in fluids*, *34*(5), 425-448.

[23] Bengio, Y., Courville, A., & Vincent, P. (2013). Representation learning: A review and new perspectives. *IEEE transactions on pattern analysis and machine intelligence*, *35*(8), 1798-1828.

[24] Chen, X., Duan, Y., Houthooft, R., Schulman, J., Sutskever, I., & Abbeel, P. (2016). Infogan: Interpretable representation learning by information maximizing generative adversarial nets. *Advances in neural information processing systems*, *29*.

[25] Tschannen, M., Bachem, O., & Lucic, M. (2018). Recent advances in autoencoder-based representation learning. *arXiv preprint arXiv:1812.05069*.

[26] Chen, J., Huang, C. K., Delgado, J. F., & Misra, S. (2023). Generating Subsurface Earth Models using Discrete Representation Learning and Deep Autoregressive Network. *arXiv preprint arXiv:2302.02594*.

[27] Misra, S., Chen, J., Falola, Y., Churilova, P., Huang, C. K., & Delgado, J. (2023, June). Massive Geomodel Compression and Rapid Geomodel Generation Using Advanced Autoencoders and Autoregressive Neural Networks. In *SPE EuropEC-Europe Energy Conference featured at the 84th EAGE Annual Conference & Exhibition*. OnePetro.

[28] Lesort, T., Díaz-Rodríguez, N., Goudou, J. F., & Filliat, D. (2018). State representation learning for control: An overview. *Neural Networks*, *108*, 379-392.

[29] Anand, A., Racah, E., Ozair, S., Bengio, Y., Côté, M. A., & Hjelm, R. D. (2019). Unsupervised state representation learning in atari. *Advances in neural information processing systems*, *32*.

[30] Shamshiri, H., & Jafarpour, B. (2010, November). Optimization of geologic CO2 storage in heterogeneous aquifers through improved sweep efficiency. In *SPE International Conference on CO2 Capture, Storage, and Utilization*. OnePetro.

[31] Cihan, A., Birkholzer, J., & Bianchi, M. (2014). Targeted pressure management during CO2 sequestration: optimization of well placement and brine extraction. *Energy Procedia*, *63*, 5325-5332.

[32] Tran, D., Shrivastava, V., Nghiem, L., & Kohse, B. (2009, October). Geomechanical risk mitigation for CO2 sequestration in saline aquifers. In *SPE Annual Technical Conference and Exhibition?* (pp. SPE-125167). SPE.

[33] Cihan, A., Birkholzer, J. T., & Bianchi, M. (2015). Optimal well placement and brine extraction for pressure management during CO2 sequestration. *International Journal of Greenhouse Gas Control*, *42*, 175-187.

[34] Zheng, F., Jahandideh, A., Jha, B., & Jafarpour, B. (2021). Geologic CO2 storage optimization

under geomechanical risk using coupled-physics models. *International Journal of Greenhouse Gas Control*, *110*, 103385.

[35] Chen, J., Gildin, E., & Killough, J. E. (2023). Physics-informed Convolutional Recurrent Surrogate Model for Reservoir Simulation with Well Controls. *arXiv preprint arXiv:2305.09056*.

[36] Chen, J., Gildin, E., & Killough, J. E. (2023). Transfer learning-based physics-informed convolutional neural network for simulating flow in porous media with time-varying controls. *arXiv preprint arXiv:2310.06319*.

[37] Kober, J., Bagnell, J. A., & Peters, J. (2013). Reinforcement learning in robotics: A survey. *The International Journal of Robotics Research*, *32*(11), 1238-1274.

[38] Kormushev, P., Calinon, S., & Caldwell, D. G. (2013). Reinforcement learning in robotics: Applications and real-world challenges. *Robotics*, *2*(3), 122-148.

[39] Kiran, B. R., Sobh, I., Talpaert, V., Mannion, P., Al Sallab, A. A., Yogamani, S., & Pérez, P. (2021). Deep reinforcement learning for autonomous driving: A survey. *IEEE Transactions on Intelligent Transportation Systems*, *23*(6), 4909-4926.

[40] Sallab, A. E., Abdou, M., Perot, E., & Yogamani, S. (2017). Deep reinforcement learning framework for autonomous driving. *arXiv preprint arXiv:1704.02532*.

[41] Ouyang, L., Wu, J., Jiang, X., Almeida, D., Wainwright, C., Mishkin, P., ... & Lowe, R. (2022). Training language models to follow instructions with human feedback. *Advances in neural information processing systems*, *35*, 27730-27744.

[42] Zhang, K., Wang, Z., Chen, G., Zhang, L., Yang, Y., Yao, C., ... & Yao, J. (2022). Training effective deep reinforcement learning agents for real-time life-cycle production optimization. *Journal of Petroleum Science and Engineering*, *208*, 109766.

[43] He, J., Tang, M., Hu, C., Tanaka, S., Wang, K., Wen, X. H., & Nasir, Y. (2022). Deep reinforcement learning for generalizable field development optimization. *SPE Journal*, *27*(01), 226-245.

[44] Nasir, Y., He, J., Hu, C., Tanaka, S., Wang, K., & Wen, X. (2021). Deep reinforcement learning for constrained field development optimization in subsurface two-phase flow. *Frontiers in Applied Mathematics and Statistics*, *7*, 689934.

[45] Sun, A. Y. (2020). Optimal carbon storage reservoir management through deep reinforcement learning. *Applied Energy*, *278*, 115660.

[46] Mnih, V., Kavukcuoglu, K., Silver, D., Rusu, A. A., Veness, J., Bellemare, M. G., ... & Hassabis, D. (2015). Human-level control through deep reinforcement learning. *nature*, *518*(7540), 529-533.

[47] Sutton, R. S., McAllester, D., Singh, S., & Mansour, Y. (1999). Policy gradient methods for reinforcement learning with function approximation. *Advances in neural information processing systems*, *12*.

[48] Mnih, V., Badia, A. P., Mirza, M., Graves, A., Lillicrap, T., Harley, T., ... & Kavukcuoglu, K. (2016, June). Asynchronous methods for deep reinforcement learning. In *International conference on machine learning* (pp. 1928-1937). PMLR.

[49] Schulman, J., Wolski, F., Dhariwal, P., Radford, A., & Klimov, O. (2017). Proximal policy optimization algorithms. *arXiv preprint arXiv:1707.06347*.

[50] Lillicrap, T. P., Hunt, J. J., Pritzel, A., Heess, N., Erez, T., Tassa, Y., ... & Wierstra, D. (2015). Continuous control with deep reinforcement learning. *arXiv preprint arXiv:1509.02971*.

[51] Haarnoja, T., Zhou, A., Abbeel, P., & Levine, S. (2018, July). Soft actor-critic: Off-policy maximum entropy deep reinforcement learning with a stochastic actor. In *International conference on*

*machine learning* (pp. 1861-1870). PMLR.

[52]  CMG. GEM, Compositional, Unconventional & Advanced Processes Simulator (2023).